\documentclass[aps,prl,twocolumn,superscriptaddress,showpacs]{revtex4}

\usepackage{graphicx}

\newcommand{\povo}{Dipartimento di Fisica, Universit\`{a} di Trento, 
and I.N.F.N., Gruppo di Trento, 
38050 Povo (TN), Italy}
\newcommand{\rthz}{\ensuremath{/\sqrt{\mathrm{Hz}}}}
\newcommand{\wsq}[1]{\ensuremath{\omega^2_{#1}}}
\newcommand{\figr}[1]{Fig. \ref{#1}}
\newcommand{\eqr}[1]{Eqn. \ref{#1}}

\begin{document}

%\preprint{}

\title{Achieving geodetic motion for LISA test masses: ground testing results}

\author{L. Carbone}
\affiliation{\povo}
\author{A. Cavalleri}
\affiliation{Centro Fisica degli Stati Aggregati, 38050 Povo (TN), Italy}
\author{R. Dolesi}
\affiliation{\povo}
\author{C. D. Hoyle}
\affiliation{\povo}
\author{M. Hueller}
\affiliation{\povo}
\author{S. Vitale}
\affiliation{\povo}
\author{W. J. Weber}
\affiliation{\povo}

\date{\today}

\begin{abstract}
The low-frequency resolution of space-based gravitational wave
observatories such as LISA (Laser Interferometry Space Antenna) hinges
on the orbital purity of a free-falling reference test mass inside a
satellite shield. We present here a torsion pendulum study of the forces
that will disturb an orbiting test mass inside a LISA capacitive
position sensor. The pendulum, with a measured torque noise floor below
10 fN$\,$m\rthz\ from 0.6 to 10 mHz, has allowed placement of
an upper limit on sensor force noise contributions, measurement of the
sensor electrostatic stiffness at the 5\% level, and detection and
compensation of stray dc electrostatic biases at the millivolt level.
\end{abstract}

%\keywords{}

\maketitle

%\section{\label{intro}Introduction}

Among the most challenging technologies needed for the LISA
gravitational wave mission is that of placing test masses in pure free
fall. Accelerations due to stray forces change the distance between
orbiting test masses, directly contaminating an interferometric
measurement of gravitational wave strain. The LISA sensitivity goal
requires acceleration noise, $a_n$, with spectral density below $3
\times 10^{-15} \, \mathrm{m/s^2}\rthz$ at frequencies down to 0.1 mHz,
or force noise of order fN\rthz\ for a $\sim$ 1~kg test mass
\cite{big_book}. 

Environmental force noise can be screened by a satellite shield
employing precision thrusters and a relative position sensor to remain
centered about the free-falling test mass. The satellite, however,
creates disturbances, particularly due to the close proximity of the
position sensor. While noise analyses have shown proposed electrostatic
sensor designs\cite{sens_LISA_symp,shoe_noise}, with 2-4 mm test mass -
sensor separations, to be compatible with the LISA goals, the low
frequencies and extreme force isolation goals require force disturbance
measurements to provide confidence in the LISA sensitivity predictions.
 
An ideal test of stray forces for LISA compares the differential noise
in the orbits of two nearby free-falling test masses. This test will be
performed, with a target acceleration noise limit of 30~fm/s$^2$\rthz\
at 1 mHz, by the LISA Test-Flight Package (LTP) \cite{LTP_LISA_symp} and
Disturbance Reduction System (DRS) \cite{DRS_LISA_symp}. In preparing
for such flight tests, we study the forces acting on a test mass that is
nearly ``free'' in a single rotational degree of freedom, suspended by a
thin torsion fiber inside a capacitive position sensor. The thermal
torque noise limit, approached in similar apparatuses \cite{wash_prd},
is several fN$\,$m\rthz\ at 1~mHz for the torsion
pendulum used here\cite{amaldi_mauro}. Dividing
by half the 40~mm test mass width, this converts to a force noise near
100~fN\rthz, within a factor 100 (10) of the LISA (LTP/DRS) force noise
target.

The translational acceleration noise relevant to LISA can be divided
into contributions from random forces $f_{str}$ acting on the test mass
(mass $m$) and from coupling to the relative motion of the satellite
(mass $M$) via any dc force gradient (or ``stiffness'' ) $k_p$. The
spacecraft motion noise arises in the position sensing noise, $x_n$, and
in the imperfect compensation of the external forces $F_{str}$ acting on
the satellite by the finite gain control loop (gain \wsq{DF}). The
residual acceleration noise $a_n$ is 
\begin{equation}
a_{n} = \frac{f_{str}}{m} + \frac{k_p}{m} 
\left( x_n + \frac{F_{str}}{M \wsq{DF}} \right)  \: \: .
\label{eqn_LISA}
\end{equation}
To characterize $a_n$, we use the torques, measured from the pendulum
twist $\phi$, acting on a LISA-like test mass inside a realistic
capacitive position sensor designed for LISA's sensing noise (spectral
density $S_{x_n}^{1/2} \sim$ nm\rthz) and electrostatic force gradient
($k_p \sim 100$ nN/m) requirements. The measured pendulum angular noise
in the LISA measurement band establishes an upper limit on the
contribution of noisy surface forces to $f_{str}$.
Measurement of the rotational stiffness due to the AC sensing
voltage characterizes a key part of the translational stiffness $k_p$.
Finally, measurement and compensation of the sensor rotational
electrostatic bias imbalance quantify and demonstrate neutralization
of a potentially important contribution of stray dc electric fields to
$f_{str}$.

\begin{figure}
 \scalebox{.8}{\includegraphics{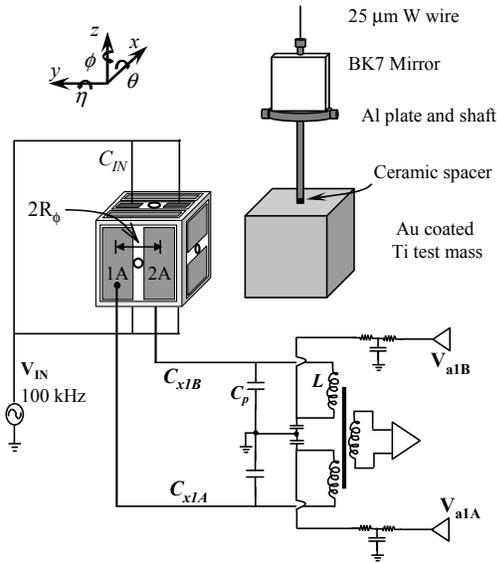}}%
\caption{\label{sens_fig_cap}Sensor electrode configuration and
circuitry, with pendulum shown at right. A 100~kHz
excitation voltage is applied capacitively through four injection
electrodes located on the sensor $z$ faces. Each of six sensing
electrode pairs is read out by a resonant inductive bridge circuit
(\cite{cqg_old}, shown here for pair \#1). The sum of the signals from
pairs 1 and 2 gives the translation in $x$, and the difference, combined
with the on-center electrode separation $2 R_{\phi} = 20.5$ mm, gives
the rotation $\phi$ about the torsion fiber axis. Actuation voltages
($V_{a1A}$ and $V_{a1B}$ for pair 1) can be applied to each electrode
through the bridge circuitry. The nominal test mass capacitances to each
injection electrode, each $x\phi$ sensing electrode, and to the entire
sensor are, respectively, $C_{IN} \approx$ 0.6~pF, $C_x \approx$~2 pF,
and $C_T \approx$~40 pF, and the excitation $V_{IN}$ =6~V$_{rms}$ AC
biases the test mass to $\alpha V_{IN} \approx$ 0.4~V$_{rms}$, where
$\alpha \equiv \frac{4 C_{IN}}{C_{T}}$.
}
\label{sens_fig}
\end{figure}

%\section{\label{exper}Experimental Apparatus}
The capacitive position sensor tested here (sketched in \figr{sens_fig}
and discussed in Refs. \cite{cqg_old,amaldi_sens}) is a variation of
that projected for LTP\cite{spie_sens,sens_LISA_symp}. With the 40 mm
test mass centered, the gap $d$ between all electrodes and the mass is 2
mm. The electrodes are gold coated Mo, and are separated from the
electrically grounded Mo housing by ceramic spacers. Differential gap
sensing measurements from six sensing electrode pairs are combined to
yield the three translational and three rotational test mass
displacements. The sensor noise floor is dominated by transformer
thermal noise, with $S_{x_n}^{1/2} \approx 0.3 \, \mathrm{nm}\rthz$ and
$S_{\phi_n}^{1/2} \approx 40 \, \mathrm{nrad}\rthz$. Integrated
actuation circuitry can apply voltages to the sensing electrodes, used
here for occasional pendulum control in $\phi$ and for electrostatic
characterization of the sensor. All electrode surfaces have a dc path to
a single circuit ground. 

The main pendulum component is the test mass itself, a hollow gold
coated Ti cube nominally 40 mm on a side, with 2 mm walls. The test mass
hangs inside the sensor, connected to the torsion fiber by an Al shaft
passing through a hole on the top $z$ face of the sensor. The shaft
supports an Al stopper plate, which limits the pendulum torsional range,
and a silvered glass mirror, which allows an independent 
autocollimator readout of the pendulum torsional ($\phi$) and swing
($\eta$) angles. The shaft assembly is grounded through the torsion
fiber and isolated from the test mass by a ceramic spacer. The total
pendulum mass is 101.4~g (the test mass weighs 80.6~g), and has a
calculated moment of inertia $I = 338 \pm 5$ g cm$^2$. 

The torsion fiber is a nominally 25 $\mu$m W wire of length 1 m. It
hangs from an upper pendulum stage with an eddy current magnetic damper,
which damps the pendulum swing mode with a 200~s decay time. The
pendulum and sensor are mounted on independent micropositioners,
allowing six degree of freedom adjustment of the relative position of
test mass inside the electrode housing. The apparatus vacuum chamber is
evacuated below 10$^{-5}$~mBar and sits in a thermally controlled room
with 50 mK long term stability.

With the sensor excitation on, the free torsional oscillation period $T_0$
is 515.1 seconds, with a quality factor $Q \approx$ 1700. The period
falls to 510.3 seconds with the sensor bias off (the period change
is due to a negative torsional sensor stiffness, to be discussed
shortly). These pendulum dynamics, combined with the calculated moment
of inertia, allow conversion of measured angular deflections in $\phi$
into torques, using the transfer function
\begin{equation}
N(\omega) = \phi(\omega) \times I \wsq{0}
\left[ 1 - \left( \frac{\omega}{\omega_0} \right) ^2 + \frac{i}{Q} \right] 
\: \: \: ,
\label{eqn_trans_func}
\end{equation}
where $\omega_0 = \frac{2 \pi}{T_0}$ and we assume a 
frequency independent pendulum loss angle $\frac{1}{Q}$.

%\section{\label{results}Measurement Results}
%\subsection{\label{force_noise}Stray Force Noise Upper Limit}

\begin{figure}
\scalebox{.9}{\includegraphics{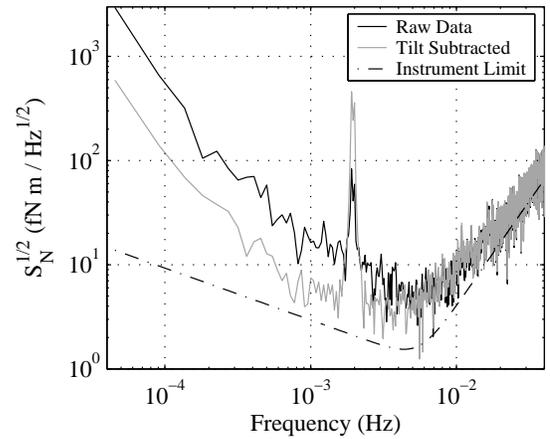}}
\caption{\label{noise_cap}Plot showing the raw (dark) and tilt
subtracted (light) pendulum torque noise, with the instrument limit
(dashed). Spectra for this
19 h measurement are calculated with a 22,000 second Hanning window,
which leaves an artificial peak near the 2 mHz pendulum resonance.}
\label{torque_noise_fig}
\end{figure}

Figure \ref{torque_noise_fig} shows typical torque noise data,
calculated from the pendulum angular noise using \eqr{eqn_trans_func}.
Also shown is the instrument limit given by the quadrature sum of
the pendulum thermal noise, $S_{N,th}^{1/2} = \sqrt{4 k_B T \frac{I
\omega_0^2}{\omega Q}}$, and the sensor readout noise,
$S_{\phi_n}^{1/2}$, converted into torque noise, which dominates above
5~mHz. Near 1~mHz, the raw torque noise is roughly 10 times the thermal
limit. We found the main excess noise source to be a coupling to
translational motion between the sensor and pendulum, due to laboratory
tilt noise. This coupling, with magnitude of order
10$^{-7}$ Nm/m, is considerably stronger than reported torsion fiber
``tilt - twist'' couplings\cite{wash_prd} and is likely related to an
electrostatic interaction involving the dielectric mirror edges. To
better characterize the sensor contribution to $f_{str}$, in the absence
of this coupling to lab motion, we calculate the instantaneous coupling
torque and subtract the resultant twist from the $\phi$ time series. The
tilt, measured in the $x$ and $y$ sensor readouts, is converted to
torque through the measured ``tilt-twist'' coupling coefficients and
then to angular twist through the pendulum transfer function
(\eqr{eqn_trans_func}). The remaining torque noise is roughly twice the
pendulum thermal noise at millihertz frequencies and below 10
fN$\,$m\rthz\ from 0.6 to 10 mHz. The excess low frequency noise is
highly correlated with the temperature fluctuations, but, to avoid
subtracting a possible sensor-related temperature effect, no further
correction has been made. 

The low noise pendulum allows precise measurement of coherently
modulated torque disturbances; we present here two such
measurements. The first concerns the electrostatic
stiffness\cite{cqg_old,amaldi_sens,spie_sens} associated with the 100
kHz sensor excitation voltage. In the approximation that all
electrode surfaces $i$ are grounded, directly or by the readout
circuitry, the test mass feels a negative electrostatic spring
proportional to the square of the bias amplitude, in rotation as well as
translation. Considering only the dc average $\left\langle V_{IN}^2
\right\rangle$, ignoring the torque at 2$\times$100 kHz, the stiffness
torque is given by 
\begin{equation}
N = \frac{\alpha^2 \left\langle V_{IN}^2 \right\rangle}{2} \! 
\left( \phi - \phi_0 \right) 
\sum_{i} \! \! \left( \frac{\partial ^2 C_i}{\partial \phi ^2} \right)
\! \equiv - \Gamma_{s} \left( \phi - \phi_0 \right)
\label{eqn_sens_stiff}
\end{equation}
Here $\phi_0$ is the (unstable) electrostatic equilibrium angle where
this torque vanishes. 

To measure the sensing stiffness $\Gamma_{s}$, we modulate the effect by
switching the 6 V$_{RMS}$ injection voltage on and off at frequency
$f_m$ = 5 mHz. The resultant pendulum deflection, measured by the
autocollimator, is proportional to $\Gamma_s$, and the amplitudes at odd
multiples of $f_m$ reflect the $1/f$ dependence of the
squarewave torque's Fourier coefficients. The measurement is made over a
range of test mass angles $\phi$ by rotating the pendulum suspension
point. Linear fitting of torque amplitude versus $\phi$ for each odd
harmonic of $f_m$ gives estimates of $\Gamma_s$ and $\phi_0$ (the latter
measured relative to the sensor zero). 

%\begin{equation}
%\Delta\Gamma (t) = 
%\Gamma_s \left( 
%\frac{1}{2} + 
%\sum_{j \: \mathrm{odd}} \frac{2}{\pi j} \sin 2 \pi j f_m t 
%\right)
%\: \: \: ,
%\label{eqn_square_wave_mod}
%\end{equation}
 
\begin{figure}
 \scalebox{.85}{\includegraphics{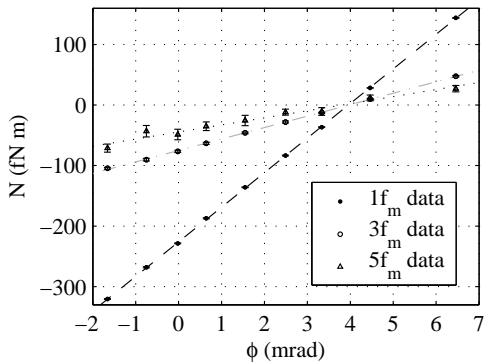}}
\caption{\label{stiff_cap}1$f_m$, 3$f_m$, and
5$f_m$ (5, 15, and 25 mHz) torque component data for the squarewave
modulated stiffness measurement. The
roughly 5:3:1 ratio of the slopes corresponds to the $1/f$
squarewave spectral content.  The sub-fN$\,$m resolution for the 1$f_m$ data 
is smaller than the point markers. }
\label{stiff_fig}
\end{figure}

Figure \ref{stiff_fig} shows modulated stiffness torques for the first
three odd harmonics of $f_m$. The resulting estimates for $\Gamma_s$ and
$\phi_0$, shown with their statistical uncertainties in Table
\ref{tab_stiff}, are in agreement for the three harmonics. Though the
statistical resolution for $\Gamma_s$ (for $1f_m$) in a given
measurement is below 1\%, repeated measurements have yielded a scatter
of $\pm 5\%$, which is currently being investigated. The measurement has
also been performed at a fixed angle for a range of $V_{IN}$, verifying
the expected $\Gamma_s \propto \left\langle V_{IN}^2 \right\rangle$
dependence. The range of $\Gamma_s$ measured with the modulation
technique is consistent with the observed period change for sensor bias
on and off, which yield $\Gamma_s = - 96 \pm 3$~fN$\,$m/mrad.
A prediction for $\Gamma_s$ obtained using a finite element capacitance
calculation\cite{diana_SPIE} gives $-96 \pm$5~fN$\,$m/mrad, with the
uncertainty dominated by machining tolerances. The measured several mrad
offsets between the torque and sensor zeros, which would coincide in a
geometrically perfect sensor, are also consistent with the machining
tolerances. 

\begin{table}
\caption{\label{tab_stiff}Electrostatic stiffness measurement results}
\begin{ruledtabular}
\begin{tabular}{c c c}
Component & $\Gamma_s$  (fN$\,$m/mrad) & $\phi_0$ (mrad) 	\\	
%\mr
1 $f_m$ & -89.76 $\pm$ 0.11 & 3.95 $\pm$ 0.01  \\ 
3 $f_m$ & -89.3 $\pm$ 0.7  & 3.97 $\pm$ 0.07   \\  
5 $f_m$ & -92 $\pm$ 6  & 3.8 $\pm$ 0.4   \\  

%\br
\end{tabular}
\end{ruledtabular}
\end{table}

Another potentially important noise source for LISA is stray dc
electrostatic fields, associated with patch or surface contamination
effects. A charged test mass feels force (torque) proportional to the
net linear (rotational) imbalances in the electrostatic potential on the
surrounding sensor surfaces. For the cosmic ray charging expected for
LISA \cite{tim_charge}, net dc potential imbalances of order 10~mV can
produce significant low frequency acceleration noise
\cite{spie_sens,pete_lowf}. To measure dc imbalances, we simulate a
charge modulation by biasing the mass with a voltage $V_{\Delta} \sin 2
\pi f_m t$ applied to the injection electrodes and measure the resulting
pendulum torque\cite{cospar_DC}.

Assigning a mean stray dc voltage $\delta V_i$ to each
conductor, the $1f_m$ torque produced in this measurement is
\begin{equation}
N_{1f_m} = - \alpha V_{\Delta} \sin 2 \pi f_m t \, \times \,
\sum_{i} \left( \frac{\partial C_i}{\partial \phi} \right) \delta V_i
\label{eqn_DC_bias}
\end{equation}
In principle, to reflect the electrostatic potential non-uniformity, the
sum over electrodes $i$ should be an integral over all surface domains
with different potentials. In the naive but useful model where each
electrode has a spatially uniform potential, the sum in
\eqr{eqn_DC_bias} reduces to 
$C_x \left( R_{\phi} / d \right) \Delta_{\phi}$,
where $\Delta_{\phi}$ is the rotational dc imbalance in the four $x\phi$
electrodes (see \figr{DC_bias_fig}) and we assume an infinite plate
model for the capacitance derivatives. 

The torque in \eqr{eqn_DC_bias}, and likewise the associated random
charge disturbance, is proportional to a net rotational dc bias
imbalance, which can be compensated by counter-biasing the $x \phi$
sensing electrodes with the actuation circuitry. Figure
\ref{DC_bias_fig} shows measured $1 f_m$ torques as a function of the
compensation voltage $V_C$. The measured torque is nulled very close to
$V_C$ = 39~mV (thus $\Delta_{\phi} \approx$ -160~mV), with balancing of
the residual $\Delta_{\phi}$ possible within 1~mV. An additional long
measurement made for $V_C$ = 39~mV has shown millivolt level stability
in the residual imbalance over a 50 h period. 

\begin{figure}
\scalebox{.9}{ \includegraphics{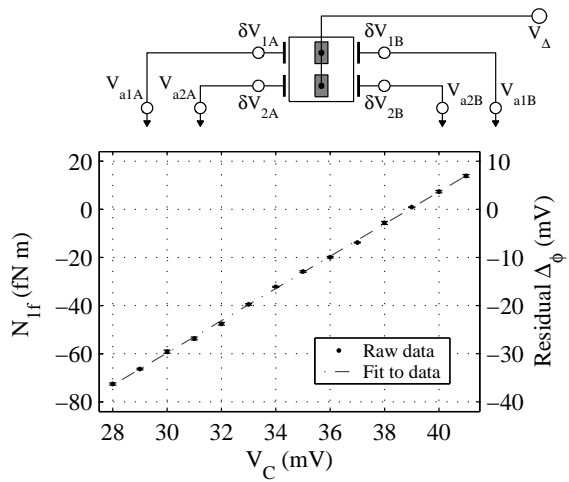}}
\caption{\label{DC_bias_fig}Plot of dc bias measurement torques for
applied bias amplitude $V_{\Delta} = 3$ V at $f_m$ = 5
mHz, as a function of the applied dc compensation voltage $V_C$. The
cartoon illustrates the measurement scheme, the average stray dc biases
(with net rotational imbalance 
$\Delta_{\phi} \equiv \delta V_{1A} + \delta V_{2B} - \delta V_{1B} - 
\delta V_{2A}$) and the applied
compensation voltages (here, we set
$V_{a1A}=V_{a2B}=-V_{a1B}=-V_{a2A}=V_{C}$). The measured torque
amplitude is proportional to a residual dc imbalance, which, for spatially
uniform stray dc biases, is $\Delta_{\phi} + 4 V_C$. 
}
\label{DC_bias_fig}
\end{figure}

%\section{\label{discussion}Discussion}

The results presented here, derived from measurements of torque on a
suspended hollow test mass that is free only along a single
rotational axis, merit some discussion concerning their representativity
of the translational forces acting on an orbiting, solid Au/Pt LISA test
mass. For the rotational stiffness measurement, $\Gamma_s$ is
proportional to the translational sensing stiffness $k_s$, with
$\partial ^2 C / \partial \phi ^2$ replaced by $ \partial ^2
C / \partial x ^2$ in \eqr{eqn_sens_stiff}. The
sensor ac bias is likely the dominant stiffness source, but it is not
the only one. The apparatus is being upgraded to measure the
full sensor-test mass coupling by modulating the sensor rotation in
$\phi$. The measurements here confirm the sensor electrostatic model and
convert to a translational stiffness uncertainty of 15~nN/m, roughly 5\%
of the total LISA stiffness budget \cite{LTP_LISA_symp}.

The rotational dc bias measurement and compensation shown here is also
directly applicable to the translational dc bias imbalance of relevance
to LISA acceleration noise. Modulating $V_{\Delta}$ also excites a
coherent force, detectable in flight \cite{cospar_DC}, proportional to
the translational imbalance, $\Delta_x$. The uniform potential model for
$\Delta_{\phi}$ is valid only as a rough number, as it neglects spatial
surface potential variation. However, the measurement itself is
sensitive to a sum over {\it all} surface domains, and thus the
compensation $V_C$ that nulls the modulated torque (or force) will also
null the torque (or force) produced by test mass charging. Compensating
a 100~mV-level imbalance to within 1~mV reduces the potentially
important random charging effect to an insignificant acceleration noise level.

The torque noise data can be cautiously converted into upper
limits on specific contributions to the force noise $f_n$ acting on LISA
test masses. The hollow test mass used here is largely immune to
gravitational or magnetic fields coupling to the bulk LISA test masses,
and the torsional mode is insensitive to net forces from important
linear temperature or field gradient effects \cite{sens_LISA_symp}. This
pendulum is designed for maximum sensitivity to surface forces, arguably
the most dangerous and unpredictable sources for 
sensors with several-mm gaps. Any electrostatic
interaction between 100~kHz circuit noise and the sensor excitation, or
between dc biases and low frequency voltage noise, produces torque noise
proportional to the net force noise. For these effects, the appropriate
arm length for converting to force noise is the electrode
half-separation $R_{\phi} = 10.25$ mm, which gives a force noise upper
limit of 1 pN\rthz\ between 0.6 and 10 mHz for such sources. Molecular
impacts give force and torque noise on all test mass faces and have a 20
mm effective arm length, yielding a 500 fN\rthz\ upper limit from 0.6 -
10 mHz, with 250~fN\rthz\ around 3~mHz. This last number corresponds,
for a solid Au/Pt test mass of the same dimensions, to acceleration
noise of 200~fm/s$^2$\rthz, a factor 7 above the LTP flight goal. 

\begin{acknowledgments}
We thank E. Adelberger and B. Hamilton for reviewing
this manuscript and D. Shaul for electrostatic analysis. 
This work was supported by ESA, INFN, and ASI. 
\end{acknowledgments}

\bibliography{pend_paper_biblio}

\begin{thebibliography}{14}
\expandafter\ifx\csname natexlab\endcsname\relax\def\natexlab#1{#1}\fi
\expandafter\ifx\csname bibnamefont\endcsname\relax
  \def\bibnamefont#1{#1}\fi
\expandafter\ifx\csname bibfnamefont\endcsname\relax
  \def\bibfnamefont#1{#1}\fi
\expandafter\ifx\csname citenamefont\endcsname\relax
  \def\citenamefont#1{#1}\fi
\expandafter\ifx\csname url\endcsname\relax
  \def\url#1{\texttt{#1}}\fi
\expandafter\ifx\csname urlprefix\endcsname\relax\def\urlprefix{URL }\fi
\providecommand{\bibinfo}[2]{#2}
\providecommand{\eprint}[2][]{\url{#2}}

\bibitem[{big()}]{big_book}
\bibinfo{note}{P. Bender et al, LISA ESA-SCI(2000)11, 2000.}

\bibitem[{\citenamefont{Dolesi et~al.}(2003)}]{sens_LISA_symp}
\bibinfo{author}{\bibfnamefont{R.}~\bibnamefont{Dolesi}} \bibnamefont{et~al.},
  \bibinfo{journal}{Class. Quant. Grav.} \textbf{\bibinfo{volume}{20}},
  \bibinfo{pages}{S99} (\bibinfo{year}{2003}).

\bibitem[{\citenamefont{Schumaker}(2003)}]{shoe_noise}
\bibinfo{author}{\bibfnamefont{B.}~\bibnamefont{Schumaker}},
  \bibinfo{journal}{Class. Quant. Grav.} \textbf{\bibinfo{volume}{20}},
  \bibinfo{pages}{S239} (\bibinfo{year}{2003}).

\bibitem[{\citenamefont{Bortoluzzi et~al.}(2003)}]{LTP_LISA_symp}
\bibinfo{author}{\bibfnamefont{D.}~\bibnamefont{Bortoluzzi}}
  \bibnamefont{et~al.}, \bibinfo{journal}{Class. Quant. Grav.}
  \textbf{\bibinfo{volume}{20}}, \bibinfo{pages}{S89} (\bibinfo{year}{2003}).

\bibitem[{\citenamefont{Hanson et~al.}(2003)}]{DRS_LISA_symp}
\bibinfo{author}{\bibfnamefont{J.}~\bibnamefont{Hanson}} \bibnamefont{et~al.},
  \bibinfo{journal}{Class. Quant. Grav.} \textbf{\bibinfo{volume}{20}},
  \bibinfo{pages}{S109} (\bibinfo{year}{2003}).

\bibitem[{\citenamefont{Smith et~al.}(1999)}]{wash_prd}
\bibinfo{author}{\bibfnamefont{G.~L.} \bibnamefont{Smith}}
  \bibnamefont{et~al.}, \bibinfo{journal}{Phys.~Rev.~D}
  \textbf{\bibinfo{volume}{61}}, \bibinfo{pages}{022001}
  (\bibinfo{year}{1999}).

\bibitem[{\citenamefont{Hueller et~al.}(2002)}]{amaldi_mauro}
\bibinfo{author}{\bibfnamefont{M.}~\bibnamefont{Hueller}} \bibnamefont{et~al.},
  \bibinfo{journal}{Class. Quant. Grav.} \textbf{\bibinfo{volume}{19}},
  \bibinfo{pages}{1757} (\bibinfo{year}{2002}).

\bibitem[{\citenamefont{Cavalleri et~al.}(2001)}]{cqg_old}
\bibinfo{author}{\bibfnamefont{A.}~\bibnamefont{Cavalleri}}
  \bibnamefont{et~al.}, \bibinfo{journal}{Class. Quant. Grav.}
  \textbf{\bibinfo{volume}{18}}, \bibinfo{pages}{4133} (\bibinfo{year}{2001}).

\bibitem[{\citenamefont{Weber et~al.}(2002{\natexlab{a}})}]{amaldi_sens}
\bibinfo{author}{\bibfnamefont{W.~J.} \bibnamefont{Weber}}
  \bibnamefont{et~al.}, \bibinfo{journal}{Class. Quant. Grav.}
  \textbf{\bibinfo{volume}{19}}, \bibinfo{pages}{1751}
  (\bibinfo{year}{2002}{\natexlab{a}}).

\bibitem[{\citenamefont{Weber et~al.}(2002{\natexlab{b}})}]{spie_sens}
\bibinfo{author}{\bibfnamefont{W.~J.} \bibnamefont{Weber}}
  \bibnamefont{et~al.}, \bibinfo{journal}{SPIE Proc.}
  \textbf{\bibinfo{volume}{4856}}, \bibinfo{pages}{31}
  (\bibinfo{year}{2002}{\natexlab{b}}).

\bibitem[{\citenamefont{Shaul and Sumner}(2002)}]{diana_SPIE}
\bibinfo{author}{\bibfnamefont{D.~N.~A.} \bibnamefont{Shaul}} \bibnamefont{and}
  \bibinfo{author}{\bibfnamefont{T.~J.} \bibnamefont{Sumner}},
  \bibinfo{journal}{SPIE Proc.} \textbf{\bibinfo{volume}{4856}},
  \bibinfo{pages}{43} (\bibinfo{year}{2002}).

\bibitem[{\citenamefont{Ara\`{u}jo et~al.}(2003)\citenamefont{Ara\`{u}jo,
  Howard, Shaul, and Sumner}}]{tim_charge}
\bibinfo{author}{\bibfnamefont{H.~M.} \bibnamefont{Ara\`{u}jo}},
  \bibinfo{author}{\bibfnamefont{A.}~\bibnamefont{Howard}},
  \bibinfo{author}{\bibfnamefont{D.~N.~A.} \bibnamefont{Shaul}},
  \bibnamefont{and} \bibinfo{author}{\bibfnamefont{T.~J.}
  \bibnamefont{Sumner}}, \bibinfo{journal}{Class. Quant. Grav.}
  \textbf{\bibinfo{volume}{20}}, \bibinfo{pages}{S311} (\bibinfo{year}{2003}).

\bibitem[{\citenamefont{Bender}(2003)}]{pete_lowf}
\bibinfo{author}{\bibfnamefont{P.~L.} \bibnamefont{Bender}},
  \bibinfo{journal}{Class. Quant. Grav.} \textbf{\bibinfo{volume}{20}},
  \bibinfo{pages}{S305} (\bibinfo{year}{2003}).

\bibitem[{cos()}]{cospar_DC}
\bibinfo{note}{W. J. Weber {\it et al.}, gr-qc/0309067 (to be published).}

\end{thebibliography}

\end{document}